\documentclass[preprintnumbers,article,amsmath,amssymb,floatfix,10pt,prd,superscriptaddress,nofootinbib,twocolumn]{revtex4}

\usepackage{doi}
\usepackage{hyperref}
\hypersetup{
  colorlinks=true,        
  linkcolor=magenta,         
  citecolor=red,         
}

\providecommand{\dif}{\mathrm{d}} \def\d{\dif}
\usepackage{graphicx}
\usepackage{dcolumn}
\usepackage{bm}
\usepackage{color}
\usepackage{enumitem}
\usepackage{amsmath}
\usepackage{amssymb}
\def\EE{{\cal E}}
\def\LL{{\cal L}}
\newcommand{\beq}{\begin{equation}}
\newcommand{\eeq}{\end{equation}}
\newcommand{\bea}{\begin{eqnarray}}
\newcommand{\eea}{\end{eqnarray}}

\usepackage{bbm}
\usepackage{amsfonts}
\usepackage{mathrsfs}
\usepackage{latexsym}
\usepackage{epsfig}
\usepackage{epstopdf}
\usepackage{epstopdf}
\usepackage{graphicx}
\usepackage{amssymb}
\usepackage{amsmath}
\usepackage{dcolumn}
\usepackage{bm}
\usepackage{color}
\usepackage{comment}
\usepackage{xcolor}

\begin{document}

\title{Testing Quantum-Corrected Black Holes with QPOs Observations: A Study of Particle Dynamics and Accretion Flow}

\author{G. Mustafa}
\email{gmustafa3828@gmail.com}
\affiliation{Department of Physics, Zhejiang Normal University, Jinhua 321004, People's Republic of China}
\affiliation{Research Center of Astrophysics and Cosmology, Khazar University, Baku, AZ1096, 41 Mehseti Street, Azerbaijan}

\author{Sushant G. Ghosh }
\email{sghosh2@jmi.ac.in}
\affiliation{Centre for Theoretical Physics, Jamia Millia Islamia, New Delhi 110025, India}
\affiliation{Astrophysics and Cosmology Research Unit, School of Mathematics, Statistics and Computer Science, University of KwaZulu-Natal, Private Bag 54001, Durban 4000, South Africa}

\author{Orhan~Donmez}
\email{orhan.donmez@aum.edu.kw}
\affiliation{College of Engineering and Technology, American University of the Middle East, Egaila 54200, Kuwait}

\author{S.K.~Maurya}
\email{ sunil@unizwa.edu.om}
\affiliation{Department of Mathematical and Physical Sciences, College of Arts and Sciences, University of Nizwa, Nizwa 616, Sultanate of Oman}

\author{Shakhzod~Orzuev}
\email{shakhzodorzuev@gmail.com}
\affiliation{New Uzbekistan University, Movarounnahr Str. 1, Tashkent 100000, Uzbekistan}
\affiliation{Institute of Fundamental and Applied Research, National Research University TIIAME, Kori Niyoziy 39, Tashkent 100000, Uzbekistan}

\author{Farruh~Atamurotov}
\email{atamurotov@yahoo.com}

\affiliation{Inha University in Tashkent, Ziyolilar 9, Tashkent, 100170, Uzbekistan}

\begin{abstract}
We study the epicyclic oscillations of test particles around rotating quantum-corrected black holes (QCBHs), characterized by mass $M$, spin $a$, and quantum deformation parameter $b$. By deriving the radial ($\Omega_r$) and vertical ($\Omega_\theta$) oscillation frequencies, we explore their dependence on spacetime parameters and show that quantum corrections ($b \neq 0$) significantly modify the dynamics compared to the classical Kerr case. Through numerical modelling of accretion around QCBHs, we further examine how $b$ influences strong-field phenomena, comparing the results with test-particle dynamics and observational data. Our analysis reveals:
 (1) Quantum corrections shift the ISCOs outward, with $b$ altering the effective potential and conditions for stable circular motion.
 (2) The curvature of the potential and thus the epicyclic frequencies change—$\Omega_r$ shows up to 25\% deviation for typical $b$ values, underscoring sensitivity to quantum effects.
 (3) Precession behavior is modified: while Lense-Thirring precession ($\Omega_{LT}$) remains primarily governed by $a$, periastron precession ($\Omega_P$) is notably affected by $b$, especially near the black hole.
 (4) Accretion disk simulations confirm the physical effects of $b$, aligning well with the test particle analysis.
Moreover, quasi-periodic oscillation (QPO) frequencies obtained via both approaches agree with observed low-frequency QPOs from sources like GRS~$1915+105$, GRO~$J1655{-}40$, XTE~$J1550{-}564$, and $H1743{-}322$. The distinct frequency profiles and altered ratios offer observational signatures that may distinguish QCBHs from classical black holes. Our findings present testable predictions for X-ray timing and a new avenue to constrain quantum gravity parameters.

\end{abstract}

\maketitle

\date{\today}


\section{Introduction} \label{ss}
Einstein’s general relativity (GR) remains a cornerstone of modern physics, providing the foundational framework for spacetime and gravitation \cite{Einstein:1915}. Its predictions—from Mercury’s perihelion precession to gravitational lensing—have been rigorously confirmed \cite{Will:2014,Abbott:2016}. By the mid-20th century, GR became essential in modeling compact objects (e.g., black holes \cite{Schwarzschild:1916}) and cosmology (e.g., cosmic expansion \cite{Riess:1998}). Despite its success, GR faces challenges in extreme regimes, motivating ongoing tests \cite{Will:2014}. Since then, it has become essential in explaining astrophysical events related to compact objects and understanding the overall structure of the universe \cite{z1,z2,z3,z4,z5,z6}.

Lately, there has been significant progress in understanding the impacts of quantum corrections on rotating black holes (BHs) \cite{q1,Afrin:2022ztr}. Different studies have proposed and examined quantum corrected (QC) versions of the Kerr BH \cite{q2}, a classical solution for rotating BHs \cite{q3}. The results indicate that quantum corrections influence observable BH properties \cite{q4}, providing new illustrations of QCBHs \cite{q5,q6,q7}. Theories like loop quantum gravity and string theory present modifications to Einstein's field equations \cite{q8}, generally through additional terms that significantly impact black hole structure near the event horizon - where quantum effects overpower. For instance, Lewandowski et al. \cite{q10} proposed a quantum-corrected black hole model that resolves the singularity problem by halting collapse at the Planck scale Studies on the characteristics of these QCBHs have examined elements such as their silhouettes, photon rings, quasi-normal modes, and gravitational lensing \cite {q11,q12,q13,q14,q15,Wang:2024iwt}.  The scarcity of rotating BH models in quantum gravity has made it difficult to test the theory using observational data. This lack of information inspired the investigation of rotating or axisymmetric extensions of spherical QCBH metrics like the one suggested by Lewandowski et al. \cite{q16,q17}. The rotating QCBH metric, derived using the Newman-Janis algorithm with parameter $\alpha$, mirrors the Kerr solution and helps test quantum gravity theories through astrophysical observations \citep{q18,q19,q20,q21,q22,Ali:2024ssf,Vachher:2024ait}.

The study of particle dynamics around black holes (BHs) is a fundamental topic in relativistic astrophysics, providing valuable insights into the underlying physical properties of information. In this context, much research has been done in this field \cite{Bardeen1972,Mashhoon1985,Stuchlik2010,Kolos2017,Oteev2018,Khan2020,Zahid2021,Khan2020arxiv,Khan2021arxiv}. The capture and motion of both massive and massless particles in parameterized BH spacetimes have been analyzed in detail, offering a deeper analysis and illustration of their behavior in strong gravitational fields. \cite{Toshmatov2021,Ahmedov2021,Abdujabbarov2020,Rahimov2024}. Also, investigations into orbital and epicyclic frequencies in axially symmetric and stationary spacetimes have tested the stability and dynamical properties of orbits in Refs. \cite{Turimov2022,Turimov2021,Turimov2023,Murodov2023,Rayimbaev2024}. Also, analytical and numerical solutions to the geodesic equations play a significant role in unveiling the rich geometric structure of BH spacetimes. The pioneering work of Hagihara ( see Ref. \cite{Hagihara1930}) laid the foundation for analytical solutions to geodesics, which have since been further tested. Grunau and Kagramanova \cite{Grunau2010} investigated the geodesics of charged particles in the Reissner–Nordström spacetime, while Chandrasekhar  in Ref. \cite{Chandrasekhar1998}, this study tested the particle trajectories around Schwarzschild, Reissner–Nordström, and Kerr BHs. Circular geodesics, in particular, have proven instrumental in analyzing quasinormal modes of BHs, as discussed by Nollert \cite{Nollert1999}. Also, the motion of electrically charged test particles in BH spacetimes has also been comprehensively studied, with various works focusing on their integrability and separation of motion equations, as shown and illustrated in Refs. \cite{Bicak1989,Stuchlik1999,Stuchlik2009,Pugliese2011}.

Epicyclic oscillations around black holes are key to understanding accretion disk dynamics and QPOs in microquasars and AGNs. Modelling the accretion disk that forms around the black holes can reveal shock waves that may develop on the disk and also allows for calculating LFQPOs and high-frequency quasi-periodic oscillations (HFQPOs). In this way, these phenomena occurring in strong gravitational fields help to identify the consistency and differences between the results predicted by alternative gravity theories, the Kerr solution, and observations \cite{Donmez2024JCAP,Donmez2024RAA,Donmez2022Univ,Donmez2014MNRAS, Koyuncu:2014MPLA,Donmez2024Universe,Donmez2024MPLA,Donmez2017MPLA,Donmez2025JHEAp,DONMEZ2024PDU}.
The parametric resonance model \cite{Kluzniak:2002bb} explains QPOs in black hole disks, while relativistic effects on frequencies in non-slender tori were analyzed in \cite{Straub:2009di}. Given their connection to test particle dynamics near the ISCO, accurately modelling QPO signals is crucial for diagnosing strong gravitational fields. The epicyclic motion of test particles characterized by orbital, radial, and latitudinal frequencies has proven helpful in explaining observed HFQPOs. These observations help constrain theoretical model parameters \cite{Bambi2012,Bambi2015,Maselli2015,Jusufi2021,Ghasemi2020,Chen2021} and references therein. Moreover, high-precision measurements from Insight-HXMT (Hard X-ray Modulation Telescope) \cite{Lu2020} and next-generation X-ray time-domain missions such as the Einstein Probe \cite{Yuan2018} are expected to place even more stringent constraints on the parameters of central compact objects.  

BHs are among the most fascinating phenomena predicted by Einstein's relativity \cite{b2,b3,b4,b5}. Alternatively, primordial BHs, which are just as fascinating, may exist because of variations in density during cosmic inflation or imperfections in the early universe. The determination of BH rotation in X-ray binary systems and the identification of merging binary BHs by LIGO provide evidence for the presence of BHs. These scenarios are justified by the revolutionary picture of the supermassive BH in galaxy M87 captured by the Event Horizon Telescope \cite{b6,b7,b8,b9}, as well as the use of radial velocity techniques to discover a star-BH binary system \cite{b10}. In addition, these results confirm the presence of BHs and offer a thorough explanation of the gravity interaction \cite{a1,a2,a3,a4,a5,a6,a7,a8}. A singularity located at the core of every BH is an area with extremely high density where the usual laws of physics cease to apply due to their intense gravitational forces. Surrounding this point of infinite gravity is the event horizon, a limit where not even light can escape the powerful pull of gravity. BHs may be distinguished by some of these parameters: \textbf{(i)} mass \cite{a9}, \textbf{(ii)} electric charge \cite{a10}, and \textbf{(iii)} angular momentum \cite{a11}. Hawking extensively explored these qualities \cite{a12,a13,a14,a15,a16}, presenting pioneering work on BH thermodynamics \cite{a17,a18,a19,a20,a21,a22} and quantum mechanics \cite{a23,a24,a25}.

The particle motion model refers to how individual particles, such as atoms, molecules, or subatomic particles, react to external forces and surroundings. Particles can illustrate different types of movement in various mediums, making this idea crucial for understanding physical systems, from basic gases to intricate quantum fields \cite{mm1,mm2,mm3,mm4,mm5,mm6}. Studying test particles dynamically in this context provides an efficient method for illustrating BH solutions in intense gravitational fields near compact objects. Another study of classical BH solutions like Schwarzschild and Kerr BHs in strong field \cite{ad17,ad18,ad19,ad20} and weak field \cite{ad21,ad22} scenarios. Nevertheless, there is significant room to explore different theories of gravity, particularly those that involve a cosmological constant and a dilaton scalar field. X-ray observations of small celestial objects have been used to investigate these alternative theories of gravity \cite{ad23,ad24,ad25}. The ISCOs are crucial for illustrating the behaviour of particles near BHs. Some studies of accretion disks constrain BH properties \cite{ad26,ad27,ad28,ad29}, while their magnetic fields significantly influence charged particles in Refs. \cite{ad30,ad31,ad32,ad33}. Another review in Refs. \cite{ad34,ad35,ad36,ad37,ad38,ad39,ad40,ad41,ad42,ad44,ad45,ad46} for explaining the interaction in the particle dynamics and spacetime around BHs.

Our study investigates the motion of test particles orbiting the rotating and axially symmetric QCBH, emphasizing how the BH's parameters influence particle motion. This quantum-corrected solution is defined by three key parameters: the BH mass $ M $, the spin parameter $ a $, and the deformation parameter $ b $, which quantifies deviations from the classical Kerr BH. Analytical formulas are obtained for the radial distributions of energy and angular momentum in stable circular orbits located in the equatorial plane. Also, by utilizing the effective potential approach, we assess the stability of these paths and examine the forces influencing the particles. The stability criteria are illuminated by determining the oscillation frequencies in the radial and latitudinal directions as dependent on $ M $, $ b $, and $ a $.
Furthermore, we investigate inner stable circular orbits, QPOs, precession effects, periastron precession (PP) and the Lense-Thirring effect. In this context, our results show that the influence of this parameter on the BH is essential to influence the movement of particles, providing a better discussion of how matter behaves near QCBHs. Additionally, we analyse QPOS effects and test how different parameters impact their frequencies and amplitudes.

This paper is organized as follows: In Section~\ref{s2}, we present the spacetime geometry of rotating quantum-corrected black holes and derive the fundamental equations governing particle motion. We analyze circular orbits, effective potentials, and the effective forces acting on test particles. Section~\ref{oscillations} investigates harmonic oscillations around stable circular orbits. We compute the fundamental frequencies ($\nu_r$, $\nu_\theta$, $\nu_\phi$) and examine periastron precession and Lense-Thirring effects, comparing results with classical Kerr black hole predictions. Finally, Section~\ref{s4} summarizes our key findings and discusses their implications for testing quantum gravity effects through astrophysical observations of black hole systems. 

Throughout this paper, unless stated otherwise, geometrized units with $G = c = 1$ are used, expressing all quantities in terms of the black hole mass. As a result, the findings are applicable to both stellar-mass and supermassive black holes.

\section{Rotating quantum corrected BH}\label{s2}
We present the spacetime geometry of a rotating black hole with quantum corrections, characterised by three fundamental parameters: the mass $M$, rotation parameter $a$, and dimensionless quantum correction parameter $b$. The line element of the rotating quantum-corrected black hole (QCBH) takes the form \cite{Ali:2024ssf,Vachher:2024ait}:
\begin{equation}\label{BH}
\begin{aligned}
ds^2 = & -\left(1 - \frac{2M(r)r}{\Sigma}\right)dt^2 - \frac{4aM(r)r\sin^2\theta}{\Sigma}dt d\phi \\
& + \frac{\Sigma}{\Delta}dr^2 + \Sigma d\theta^2 + \left(r^2 + a^2 + \frac{2a^2M(r)r\sin^2\theta}{\Sigma}\right) \\
& \times \sin^2\theta d\phi^2,
\end{aligned}
\end{equation}

where 
\bea
\Delta &&=r^2+a^2-2 M(r) r, \quad M(r)=M-\frac{b M^4}{2 r^3},\nonumber\\
\Sigma &&=r^2+a^2 \cos ^2 \theta.\nonumber
\eea
The metric (\ref{BH}) contains three physical parameters: the mass $M$ sets the overall scale of the black hole, the rotation parameter $a$ (with $0 \leq a \leq M$) determines the angular momentum, and the quantum correction parameter $b$ quantifies deviations from the Kerr black hole. Several critical limits emerge from this solution: When $b = 0$, the metric reduces exactly to the Kerr solution. We recover the Schwarzschild metric for $a = 0$ and $b = 0$. The case $a = 0$ with $b \neq 0$ corresponds to a spherically symmetric quantum-corrected black hole.
The horizon locations are  the roots of the equation $\Delta(r) = 0$, which becomes:
\begin{equation}
r^2 + a^2 - 2\left(M - \frac{bM^4}{2r^3}\right)r = 0.
\end{equation}
This quintic equation governs inner and outer horizons, with the outer solution representing the event horizon. The quantum correction parameter $b$ modifies the horizon radii compared to the classical Kerr case, with the correction term becoming significant at small radii where quantum effects are expected to dominate.

\subsection{Orbital Dynamics Around Rotating QCBHs} 
The motion of neutral particles in the quantum-corrected spacetime is governed by the Hamiltonian formulation \cite{BZ1, BZ2, BZ3}. The relativistic Hamiltonian takes the form:
\begin{equation}
H = \frac{1}{2}g^{\alpha\beta}p_{\alpha}p_{\beta} + \frac{1}{2}m^2,
\end{equation}
where $m$ is the particle mass, $p^{\gamma} = mu^{\gamma}$ the four-momentum, $u^{\gamma} = dx^{\gamma}/d\tau$ the four-velocity, and $\tau$ the proper time. The corresponding Hamilton's equations \cite{BZ1, BZ2, BZ3} are:
\begin{equation}
\frac{dx^{\gamma}}{d\zeta} = mu^{\gamma} = \frac{\partial H}{\partial p_{\gamma}}, \quad
\frac{dp_{\gamma}}{d\zeta} = -\frac{\partial H}{\partial x^{\gamma}},
\end{equation}
with $\zeta = \tau/m$ being an affine parameter. The spacetime symmetries yield two conserved quantities \cite{BZ1, BZ2, BZ3}:
\begin{equation}
\frac{p_t}{m} = g_{tt}u^{t} + g_{t\phi}u^{\phi} = -\mathcal{E}, \quad
\frac{p_{\phi}}{m} = g_{\phi\phi}u^{\phi} + g_{t\phi}u^{t} = \mathcal{L},
\end{equation}
where $\mathcal{E} = E/m$ denotes the specific energy, which is a dimensionless quantity in geometrized units, and $\mathcal{L} = L/m$ is the specific angular momentum of the test particle. For the considered rotating QCBH, the Hamiltonian takes the following form:
\begin{eqnarray}
H=\frac{b^2 p_{r}^{2}+b H_2 r^2+H_3 r^4}{2 r^4 \left(b+H_1 r^2\right)},
\end{eqnarray}
where
\begin{eqnarray*}
H_1&&=a^2+(r-2) r,\\
H_2&&=a^2 \left(\mathcal{E}^2+2 p_{r}^{2}\right)-2 a \mathcal{E} \mathcal{L}+\mathcal{L}^2\\&&+p_{\theta }^{2}+2 p_{r}^{2} r^2-4,
p_{r}^{2} r+r^2,\\
H_3&&=a^4 p_{r}^{2}+a^2 \left(r \left(-\left(\mathcal{E}^2 (r+2)\right)+2 p_{r}^{2} (r-2)+r\right)\right.\\&&+\left.p_{\theta }^{2}\right)+4 a \mathcal{E} \mathcal{L} r+r \left(r \left(r \left(\mathcal{E}^2 (-r)+r-2\right)\right.\right.\\&&+\left.\left.p_{r}^{2} (r-2)^2\right)+\mathcal{L}^2 (r-2)+p_{\theta}^{2} (r-2)\right)
\end{eqnarray*}
The specific energy and specific angular momentum of particles around rotating QCBHs are calculated as:
\begin{eqnarray}
\label{r19}
\mathcal{E}&&=\frac{a^2 \left(2 b r-r^4\right)-\mathcal{E}_3 \mathcal{E}_2 a+b r^3+(r-2) r^6}{\sqrt{\mathcal{E}_1} r^2 \left(a^2 \left(2 b-r^3\right)+r^6\right)},\\\label{r20}
\mathcal{L}&&=\frac{a^3 \left(2 b r-r^4\right)+3 a r^3 \left(b-r^3\right)+\mathcal{E}_2 r^6+\mathcal{E}_5}{\sqrt{\mathcal{E}_1} r^2 \left(a^2 \left(2 b-r^3\right)+r^6\right)},
\end{eqnarray}

\begin{eqnarray*}
\mathcal{E}_1&&=\frac{2 a^3 r \left(r^3-2 b\right)^{3/2}+3 b r^6+(r-3) r^9+\mathcal{E}_4}{\left(a^2 \left(2 b-r^3\right)+r^6\right)^2},\\
\mathcal{E}_2&&=\sqrt{r^3-2 b}, \;\;\mathcal{E}_3=b-2 r^3,\\
\mathcal{E}_4&&=6 \mathcal{E}_2 a r^3 \left(r^3-b\right)-3 a^2 \left(2 b^2-b (2 r+3) r^3\right.\\&&+\left.(r+1) r^6\right),\;\;\mathcal{E}_5=\mathcal{E}_2 a^2 \left(r^3 (r+2)-b\right)
\end{eqnarray*}

\begin{figure*}
\centering 
\includegraphics[scale=0.68]{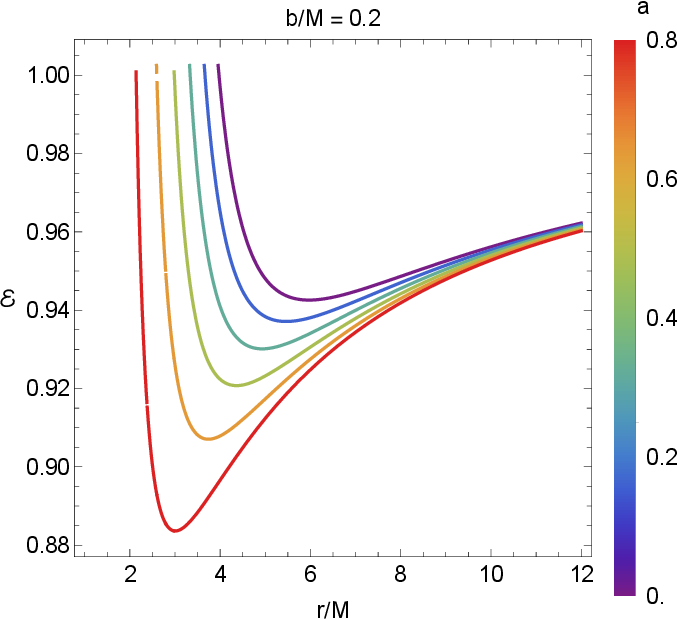}\;\;\;\;\;\;
\includegraphics[scale=0.68]{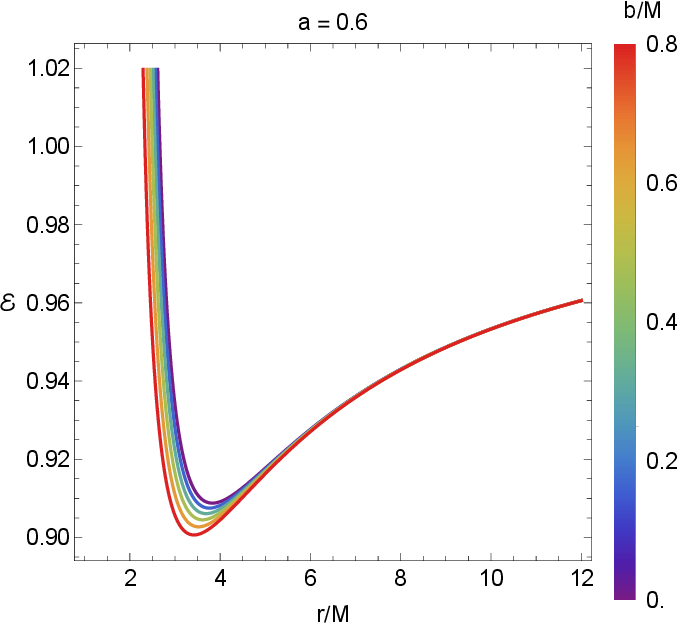}
\caption{Graphical evolution of energy of particles encircling the rotating QCBHs.}\label{Fig1}
\label{figENG}
\end{figure*}

\begin{figure*}
\centering 
\includegraphics[scale=0.68]{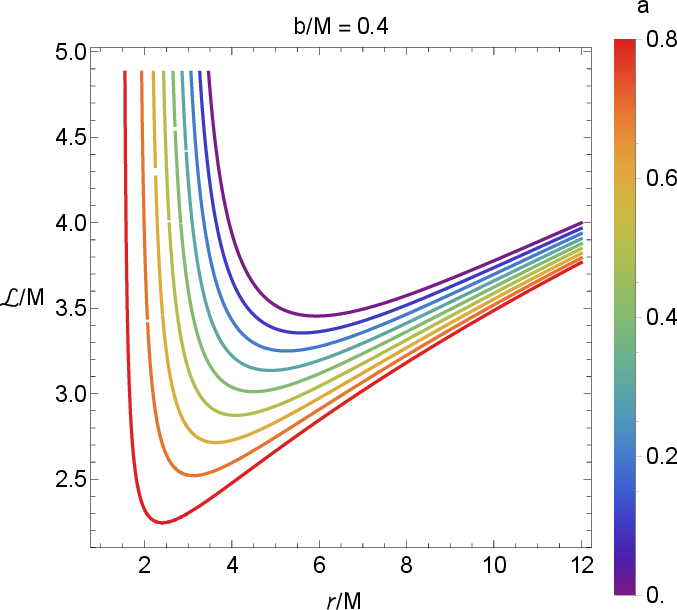}\;\;\;\;\;\;
\includegraphics[scale=0.68]{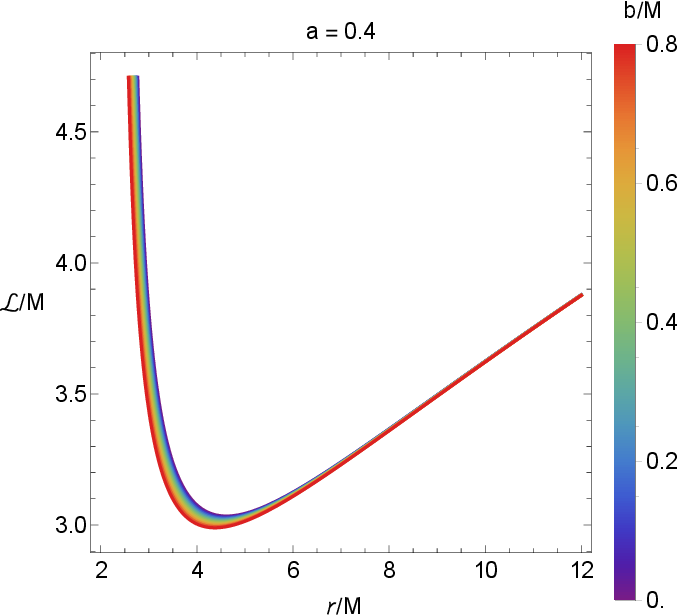}
\caption{Graphical evolution of angular momentum of particles encircling the rotating QCBHs.} \label{Fig2}
\label{fig_ANG}
\end{figure*}

The orbital energy characteristics of test particles around the rotating QCBH are depicted in Fig.~\ref{figENG}. The left panels demonstrate the dependence on the spin parameter $a$, revealing that particle energy decreases monotonically with increasing $a$ for fixed radial coordinate $r$. This behaviour reflects the enhanced frame-dragging effects in rapidly rotating black holes. The right panels illustrate the influence of the quantum correction parameter $b$, showing that larger quantum corrections correspond to systematically higher orbital energies than the classical Kerr case ($b=0$). Notably, the energy difference between QCBH and Kerr solutions becomes most pronounced in the strong-field region ($r \lesssim 10M$), suggesting that quantum corrections significantly modify the potential well structure near the event horizon.

A comparative analysis reveals two key features: For given $(r, a)$ values, quantum-corrected spacetimes consistently yield higher particle energies than their classical Kerr counterparts. The energy reduction rate with increasing $a$ is steeper for larger values of $b$, indicating an interplay between rotational and quantum-gravitational effects. The angular momentum distribution, shown in Fig.~\ref{fig_ANG}, exhibits complementary behaviour. The left panels demonstrate that increasing the spin parameter $a$ leads to lower specific angular momentum $\mathcal{L}$ at a fixed radius, with the reduction being most dramatic for $r \lesssim 6M$. The right panels reveal that quantum corrections (parameterized by $b$) similarly suppress the required angular momentum for circular orbits, though with characteristically different radial dependence. Three crucial trends emerge: $a$ and $b$ parameters show an inverse relationship with $\mathcal{L}$, with the spin parameter $a$ having more pronounced effects in the strong-field regime. The angular momentum increases monotonically with radial distance $r$, approaching the classical Keplerian limit ($\mathcal{L} \propto \sqrt{Mr}$) at large distances. Quantum-corrected spacetimes require systematically higher angular momentum than Kerr black holes at equivalent radii, particularly in the $3M < r < 10 M$ range. 

These energy and angular momentum characteristics directly affect accretion disk physics, as they influence the location of the ISCO, the efficiency of energy extraction mechanisms, and the spectral signatures of disk emission. 
The observed deviations from Kerr predictions suggest potentially observable signatures of quantum corrections in astrophysical black hole systems.
\begin{figure*}
\centering 
\includegraphics[scale=0.68]{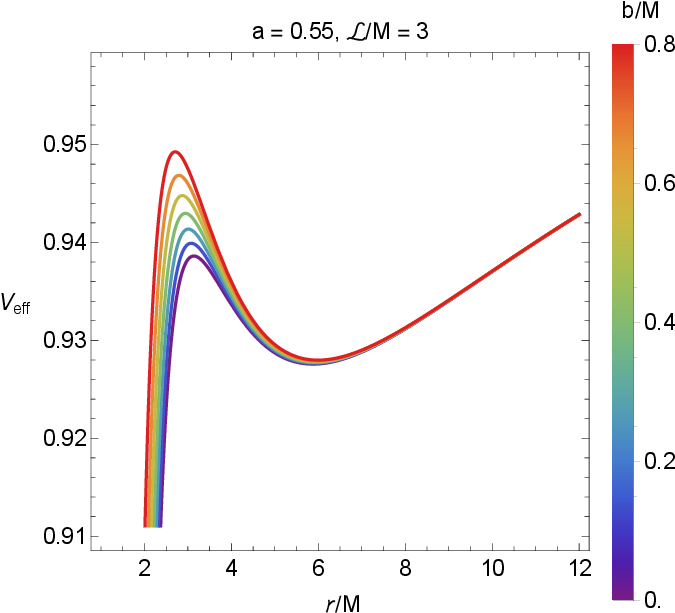}\;\;\;\;\;\;
\includegraphics[scale=0.68]{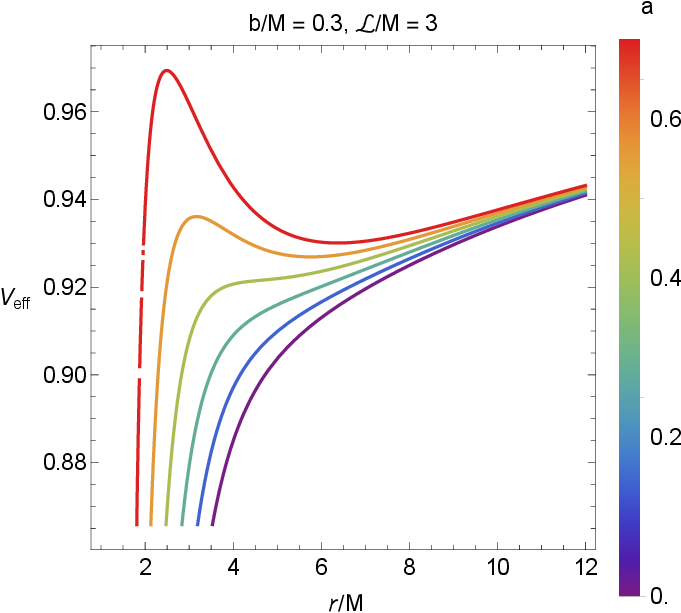}
\caption{Graphical evolution of the effective potential of particles around rotating QCBHs.
}\label{fig_eff}
\end{figure*}

\subsection{Effective potential}
The effective potential formalism for particle motion in curved spacetime arises naturally from the normalization condition $g_{\nu\sigma} u^\nu u^\sigma = -1$ \cite{MTW1973, Chandrasekhar1983}. This yields the general expression:
\begin{equation}
V_{\text{eff}}(r, \theta) = g_{rr}\, \dot{r}^2 + g_{\theta \theta}\, \dot{\theta}^2,
\end{equation}
where $\dot{r} = dr/d\tau$ and $\dot{\theta} = d\theta/d\tau$ are proper time derivatives. The potential can be expressed in terms of metric components and constants of motion as \cite{Bardeen1972, Wilkins1972}:
\begin{equation}
V_{\text{eff}}(r, \theta) = \frac{\EE^2 g_{\phi \phi} + 2 \EE \LL g_{t \phi} + \LL^2 g_{tt}}{g_{t\phi}^{2} - g_{tt} g_{\phi\phi}} - 1.
\end{equation}

For the spacetime geometry considered here, the explicit form becomes \cite{Quevedo2011, Kato1990}:
\begin{equation}
\label{eq:Veff_main}
V_{\text{eff}}(r, \theta) = \frac{a^2 r^3 \sqrt{V_1} - a \mathcal{L} \left(b - 2 r^3\right) + r \sqrt{V_1} V_2}{a^2 \left(r^3 (r+2) - b\right) + r^6},
\end{equation}
with the auxiliary functions:
\begin{align}
V_1 &= \frac{a^2 \left(r^3 (r+2) - b\right) + r^4 \left(\mathcal{L}^2 + r^2\right)}{r^2 \left(a^2 + (r-2) r\right) + b}, \label{eq:V1} \\
V_2 &= b + (r-2) r^3. \label{eq:V2}
\end{align}

This potential is beneficial for analysing bound orbits and stability conditions without solving the full equations of motion \cite{Ryan1995, Levin2009}. In the equatorial plane ($\theta = \pi/2$), circular orbits satisfy:
\begin{equation}
V_{\text{eff}}(r) = 0, \quad \frac{\partial V_{\text{eff}}(r)}{\partial r} = 0. \label{eq:circular_conditions}
\end{equation}
These conditions are fundamental for determining the ISCO and other characteristic radii \cite{Abramowicz2013}.
The extrema of the effective potential $V_{\text{eff}}(r)$ identify circular orbits, with minima corresponding to stable configurations and maxima to unstable ones. While Newtonian gravity predicts a fixed ISCO radius independent of angular momentum $\mathcal{L}$, relativistic systems exhibit $\mathcal{L}$-dependent orbital positions due to additional degrees of freedom \cite{Bardeen1972}. 

For Schwarzschild black holes, the effective potential displays two extremal points for each angular momentum value, as shown in Fig.~\ref{fig_eff}. The ISCO location emerges from solving the simultaneous conditions:
\begin{equation}
\frac{\partial V_{\text{eff}}}{\partial r} = 0, \quad 
\frac{\partial^2 V_{\text{eff}}}{\partial r^2} \geq 0.
\label{eq:isco_conditions}
\end{equation}

The orbital stability depends critically on the spin parameter $a$ and deformation parameter $b$. As either parameter increases, the region of stable orbits contracts significantly. Notably, quantum-corrected black holes (QCBHs) develop shallower potential minima near the horizon compared to their Kerr counterparts at identical spins \cite{Quevedo2011}, with Kerr solutions exhibiting $V_{\text{eff}}$ minima at smaller radii. This behaviour reflects how quantum corrections ($b \neq 0$) modify the spacetime geometry near the event horizon.

\subsection{Effective force} 

\begin{figure*}
\centering 
\includegraphics[scale=0.68]{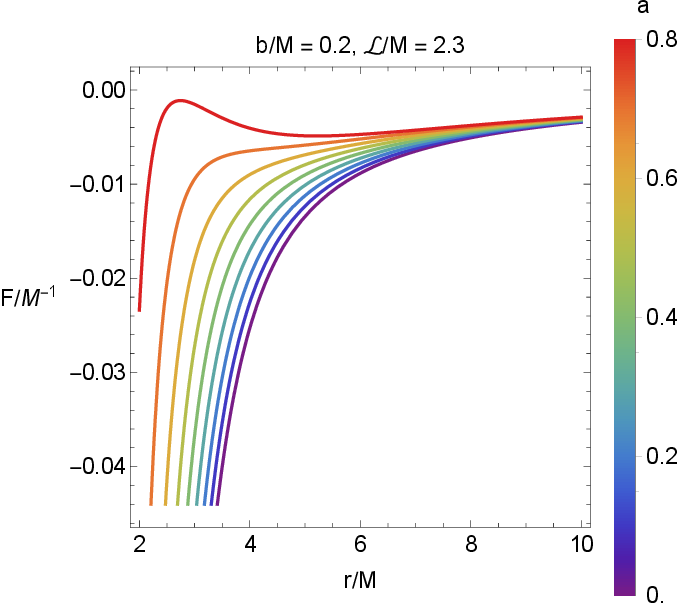}\;\;\;\;\;\;
\includegraphics[scale=0.68]{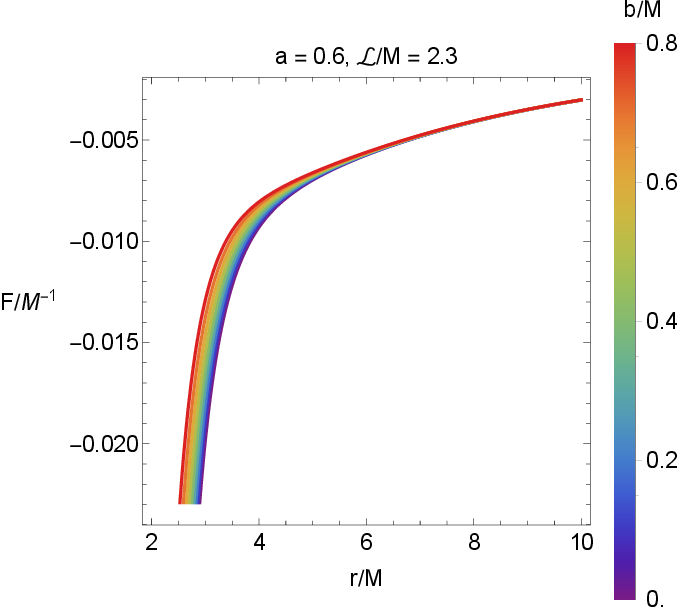}
\caption{Graphical evolution of effective force for particles moving around rotating QCBHs.}\label{figForce}
\end{figure*}

The radial effective force governing particle dynamics,
\begin{equation}
F = -\frac{1}{2}\frac{dV_{\text{eff}}}{dr},
\end{equation}
exhibits strong dependence on both the quantum correction parameter $b$ and spin parameter $a$, as shown in Fig.~\ref{figForce}. The force magnitude increases monotonically with both parameters, transitioning from weakly attractive at small $a,b$ to strongly repulsive at larger values. Notably, quantum-corrected black holes generate consistently stronger forces than their Kerr counterparts with equivalent spin parameters, revealing how quantum gravitational effects modify the near-horizon geometry.

\begin{figure*}
\centering 
\includegraphics[width=\hsize]{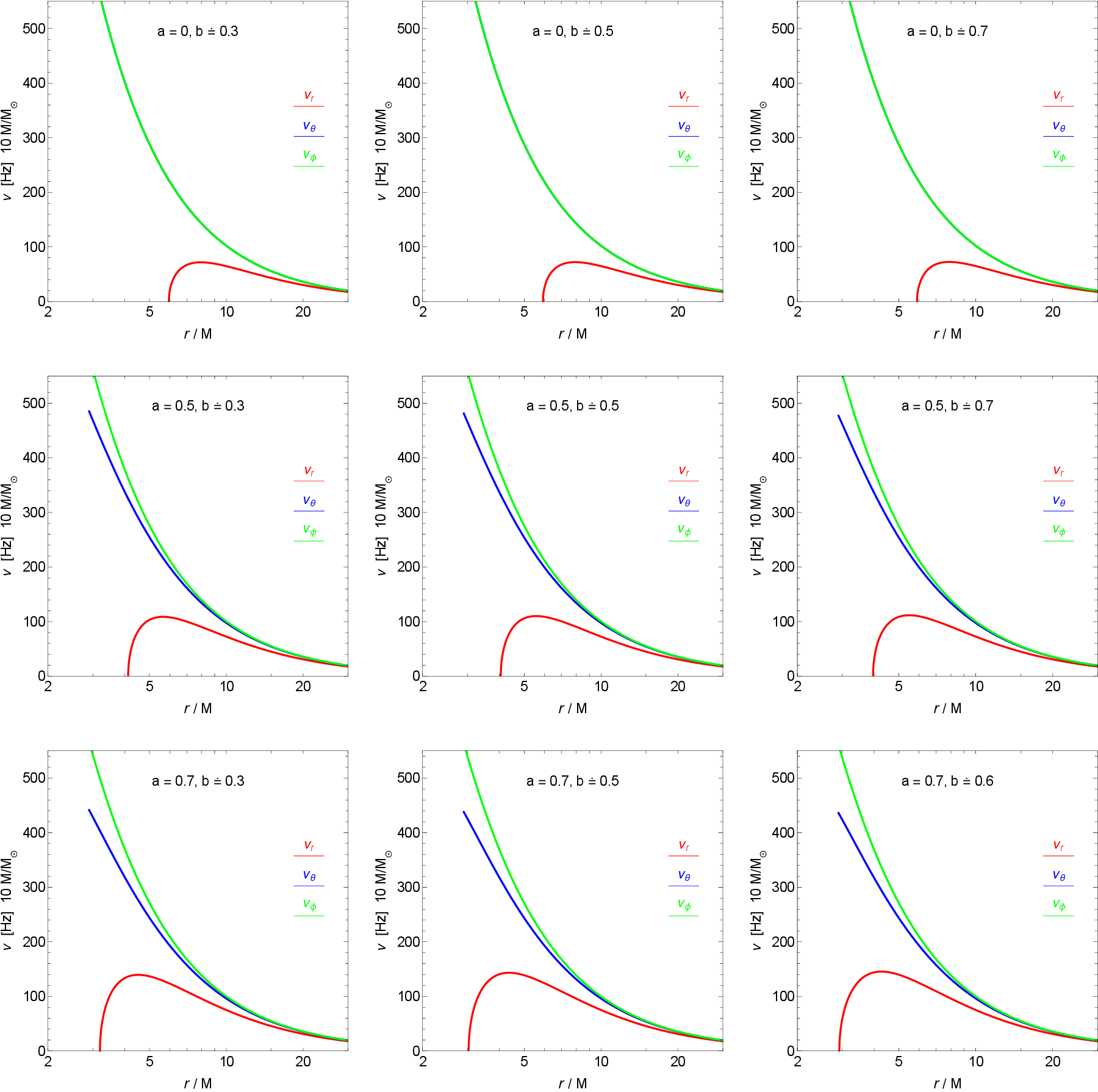}
\caption{Graphical evolution of fundamental frequencies of particles moving around rotating QCBHs.
}\label{figFRQ}
\end{figure*}

\begin{figure*}
\centering 
\includegraphics[scale=0.68]{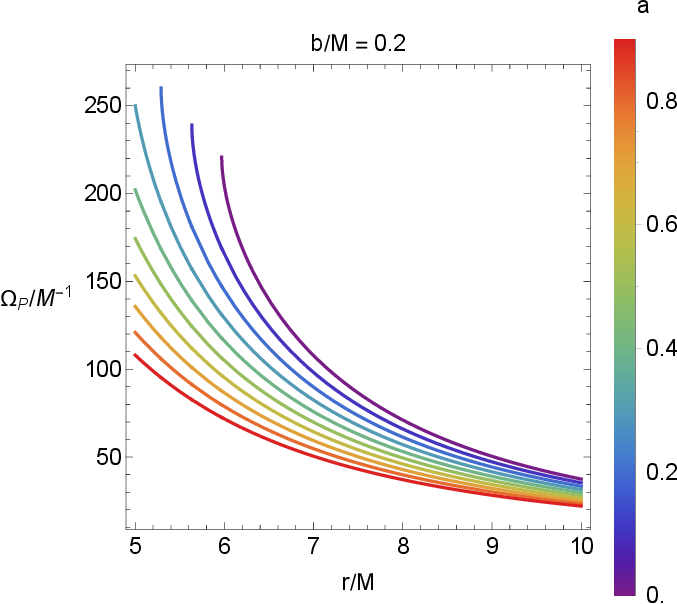}\;\;\;\;\;\;
\includegraphics[scale=0.68]{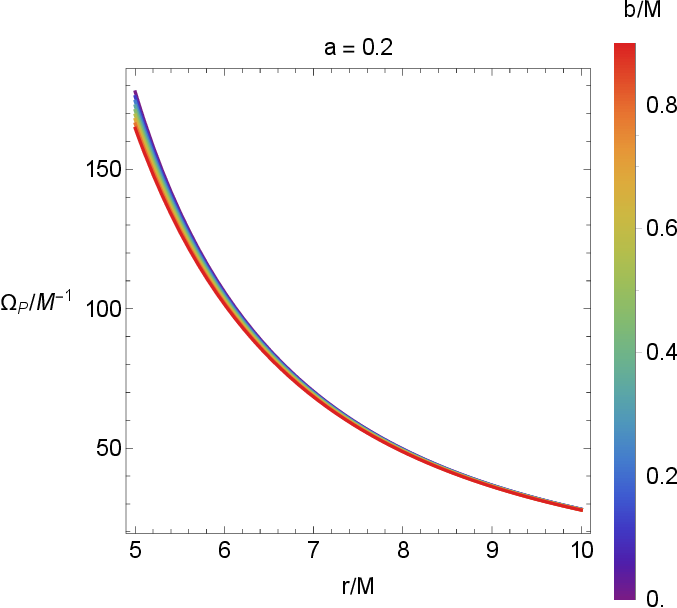}
\caption{Graphical evolution of periastron frequency of particles around rotating QCBHs.}\label{fig_periastron}
\end{figure*}

\begin{figure*}
\centering 
\includegraphics[scale=0.68]{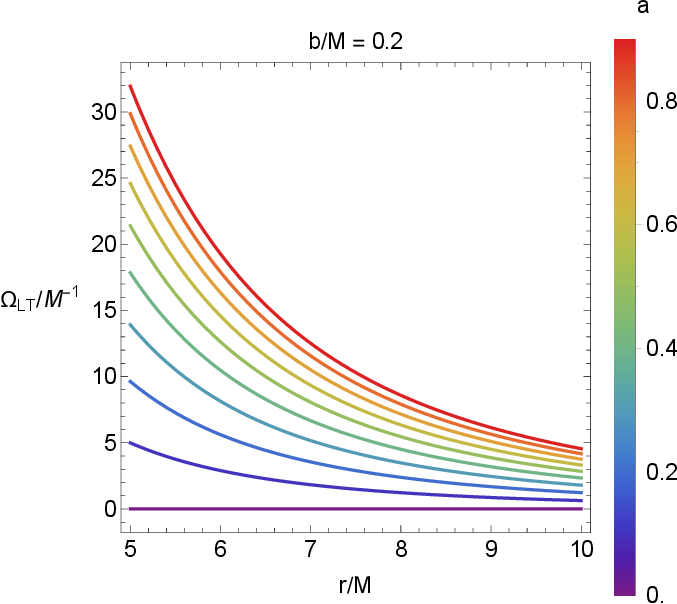}\;\;\;\;\;\;
\includegraphics[scale=0.68]{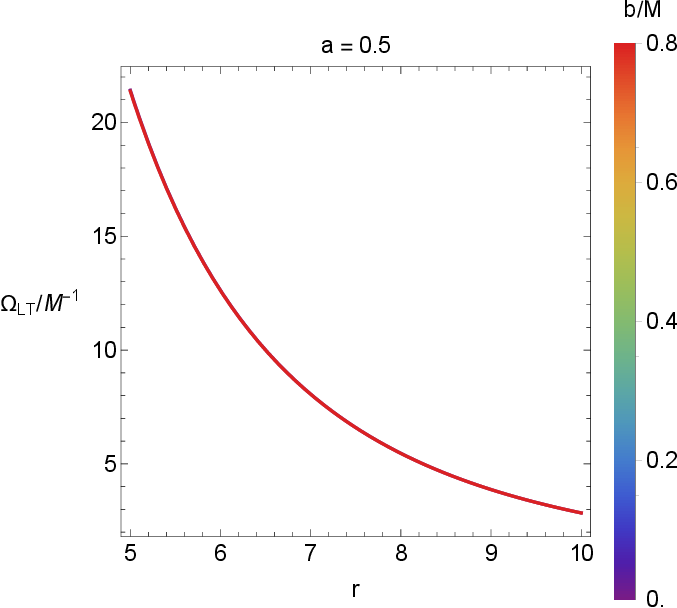}
\caption{Graphical evolution of Lense-Thirring precession frequency around rotating QCBHs.}\label{fig_LT}
\end{figure*}

\section{Analyzing harmonic oscillations in the context of circular orbital motion} \label{oscillations}

We consider small perturbations about stable circular orbits in the equatorial plane to analyze the oscillatory dynamics of neutral particles. When displaced from equilibrium, a test particle executes epicyclic motion, a superposition of radial and vertical harmonic oscillations. The fundamental frequencies of these oscillations, as measured by a comoving observer, are determined by the curvature of the effective potential and given by \cite{BZ1,BZ2,BZ3}
\bea\label{Freq-2}
\omega_{r}^{2} &=&  \frac{-1}{2\, g_{rr}} \frac{\partial^{2} V_{\rm eff} (r, \theta)}{\partial r^{2}},\\\label{Freq-3}
\omega_{\theta}^{2} &=& \frac{-1}{2\, g_{\theta \theta}} \frac{\partial^{2} V_{\rm eff}(r, \theta)}{\partial \theta^{2}},\\\label{Freq-4}
\omega_\phi &=& \frac{\d \phi}{\d \tau}.
\eea
Analyzing fundamental frequencies ($\omega_r$, $\omega_\theta$, $\omega_\phi$) and their ratios reveal crucial insights into the geometry of epicyclic motions around stable circular orbits. In Newtonian gravity, the frequency degeneracy ($\omega_r = \omega_\theta = \omega_\phi$) produces simple elliptical orbits around spherical masses. However, Schwarzschild black holes break this degeneracy, establishing the hierarchy $\omega_r < \omega_\theta = \omega_\phi$. This frequency splitting generates two distinct relativistic effects: (1) a periapsis advance governed by $\omega_\phi - \omega_r$ and (2) an orbital plane precession determined by $\omega_\phi - \omega_\theta$, which becomes increasingly pronounced as particles approach the black hole.

\subsection{Frequencies measured by far observer} 

The locally measured angular frequencies $\omega_\alpha$ (given in Eqs.~\ref{Freq-2}--\ref{Freq-4}) relate to those observed at infinity $\Omega_\alpha$ through the redshift transformation \cite{BZ1,BZ2,BZ3}
\begin{equation}\label{frequencies}
\Omega_{\alpha} = \omega_{\alpha} \left(\frac{d\tau}{dt}\right),
\end{equation}
where the redshift factor is determined by the spacetime metric:
\begin{equation}\label{redshift}
\frac{dt}{d\tau} = -\frac{E g_{\phi\phi} + L g_{t\phi}}{g_{tt}g_{\phi\phi} - g_{t\phi}^2}.
\end{equation}
For a distant observer measuring oscillation frequencies in physical units, we convert to dimensionless form using the characteristic scale $c^3/GM$, yielding:
\begin{equation}\label{nu_rel}
\nu_j = \frac{1}{2\pi}\left(\frac{c^3}{GM}\right) \Omega_j \quad [\mathrm{Hz}].
\end{equation}
Here, $ j \in \{r, \theta, \phi\} $, and $ \Omega_r $, $ \Omega_\theta $, and $ \Omega_\phi $ denote angular frequencies, as observed by a distant observer, for the radial, latitudinal, and axial components, respectively. For a rotating QCBH, the expressions for $ \Omega_{\alpha} $ are given by:
\bea\label{nu_r}
\Omega_{r}^{2} &=&\frac{a^4 \Omega _3+a^2 \Omega _2-2 a \mathcal{E} \mathcal{L} \Omega _4+3 b^3 \mathcal{L}^2+\Omega _5}{\Omega _1}
,\\\label{nu_theta}
\Omega_{\theta}^{2} &=&\frac{\Omega _6 \left(b+H_1 r^2\right)}{r^2 \left(a^2 \mathcal{E} \left(r^3 (r+2)-b\right)+\mathcal{E}_3 a \mathcal{L}+\mathcal{E} r^6\right){}^2}
,\\\label{nu_phi}
\Omega_\phi &=& \frac{2 a b-a r^3+\mathcal{E}_2 r^3}{a^2 \left(2 b-r^3\right)+r^6},
\eea

where
\begin{eqnarray*}
\Omega _1&&=r^4 \left(a^2 \mathcal{E} \left(r^3 (r+2)-b\right)+\mathcal{E}_3 a \mathcal{L}+\mathcal{E} r^6\right){}^2,\\
\Omega _2&&=3 b^3 \mathcal{E}^2+b^2 r^2 \left(\mathcal{E}^2 r (7 r-18)+9 \mathcal{L}^2\right)+2 b r^5 \\&&\times \left(\mathcal{E}^2 r \left(15 r^2-32 r+18\right)+3 \mathcal{L}^2 (4 r-7)\right)\\&&-\left(r^8 \left(2 \mathcal{E}^2 r \left(3 r^2-14 r+12\right)+\mathcal{L}^2 \left(r^2+6 r-12\right)\right)\right),\\
\Omega _3&&=9 b^2 \mathcal{E}^2 r^2+2 b r^4 \left(3 \mathcal{E}^2 r (5 r-7)+5 \mathcal{L}^2\right)-2 r^7\\&& \times \left(3 \mathcal{E}^2 (r-2) r+\mathcal{L}^2\right),\\
\Omega _4&&=3 b^3+2 b^2 (4 r-9) r^3+b \left(21 r^2-50 r+36\right) r^6-4 \\&&\times\left(3 r^2-8 r+6\right) r^9,
\end{eqnarray*}

\begin{eqnarray*}
\Omega _5&&=-2 a^6 \mathcal{E}^2 r^4 \left(r^3-5 b\right)+4 a^5 \mathcal{E} \mathcal{L} r^4 \left(r^3-5 b\right)-6 a^3 \\&&\times \mathcal{E} \mathcal{L} r^2 \left(3 b^2+b (9 r-14) r^3-2 (r-2) r^6\right)-6 b^2 \mathcal{E}^2 r^6\\&&+9 b^2 \mathcal{L}^2 (r-2) r^3+2 b \mathcal{E}^2 (5 r-3) r^9+9 b \mathcal{L}^2 (r-2)^2 r^6\\&&+r^9 \left(3 \mathcal{L}^2 (r-2)^3-2 \mathcal{E}^2 r^4\right),\\
\Omega _6&&=-\mathcal{E}_3 a^4 \mathcal{E}^2+2 \mathcal{E}_3 a^3 \mathcal{E} \mathcal{L}+a^2 \left(2 r^3-b\right)\\&&\times \left(\mathcal{E}^2 r^2+\mathcal{L}^2\right)+\mathcal{L}^2 r^2 V_2
\end{eqnarray*}
Figure~\ref{figFRQ} presents the radial profiles of epicyclic frequencies $\nu_j$ for neutral particles orbiting a rotating QCBH, demonstrating their dependence on the spin parameter $a$ and quantum correction $b$ as measured by a distant observer. In the non-rotating case ($a=0$), the radial and vertical frequency profiles degenerate, as evident in the plots' top row. Introducing rotation ($a>0$) or quantum corrections ($b>0$) breaks this degeneracy. Both parameters cause the radial frequency peak to shift inward toward smaller radii, with the effect being more pronounced for QCBHs than for Kerr black holes with identical spin. This inward migration of frequency profiles reveals how quantum corrections enhance relativistic effects near the horizon, modifying the orbital dynamics compared to classical Kerr predictions.

\subsection{Periapsis and Lense-Thirring in Quantum-Corrected Spacetime} 
We analyze the Lense-Thirring precession (LTP) and periapsis precession for neutral test particles in slightly inclined orbits ($\theta \approx \pi/2$) around a rotating quantum-corrected black hole (QCBH). The particle's epicyclic motion with radial frequency $\Omega_r$ enables precise measurement of these relativistic effects.

The precession frequencies are defined as \cite{BZ1,BZ2,BZ3}:
\begin{align}
\Omega_P &= \Omega_\phi - \Omega_r \quad \text{(periapsis precession)} \label{eq:peri} \\
\Omega_{LT} &= \Omega_\phi - \Omega_\theta \quad \text{(Lense-Thirring precession)} \label{eq:lt}
\end{align}

These effects fundamentally differ from Newtonian gravity, where $\Omega_r = \Omega_\phi$. The periapsis precession $\Omega_P$, visualized in Fig.~\ref{fig_periastron}, describes the in-plane orbital advance, while $\Omega_{LT}$ (nonzero only when $a\neq0$) quantifies the nodal plane precession from frame-dragging. 

For non-rotating black holes ($a=0$), $\Omega_\theta = \Omega_\phi$ eliminates LTP while preserving periapsis precession. As shown in Fig.~\ref{fig_LT}, increasing the spin parameter $a$ produces two competing effects: it enhances $\Omega_{LT}$ through stronger frame drag while reducing $\Omega_P$. The quantum correction parameter $b$ shows a markedly different influence - while it negligibly affects $\Omega_{LT}$, it significantly modifies the $\Omega_P$ profiles compared to classical Kerr solutions.


\section{Comparison of Test Particle Dynamics and Numerical Accretion Flow Around QCBH} \label{orhan2}

Here, we compare the motion of the test particle in the strong gravitational field around the QCBH with the behaviour of the accretion disk formed in the same strong gravitational region, i.e., very close to the horizon of the QCBH. By doing so, we analyze the physical implications introduced by the $b$ in each case. To achieve this, we numerically solve the general relativistic hydrodynamic equations (GRH), modelling the matter infall through Bondi-Hoyle-Lyttleton (BHL) accretion toward the black hole, and thereby simulate the formation of the accretion disk and the resulting shock cone in the equatorial plane. The boundary conditions, initial conditions, and other physical parameters used in the modelling are described in detail in \cite{DONMEZ2024PDU,Donmez2022Univ,Donmez2024MPLA}.

\begin{figure*}
  \vspace{1cm}
  \center
  \psfig{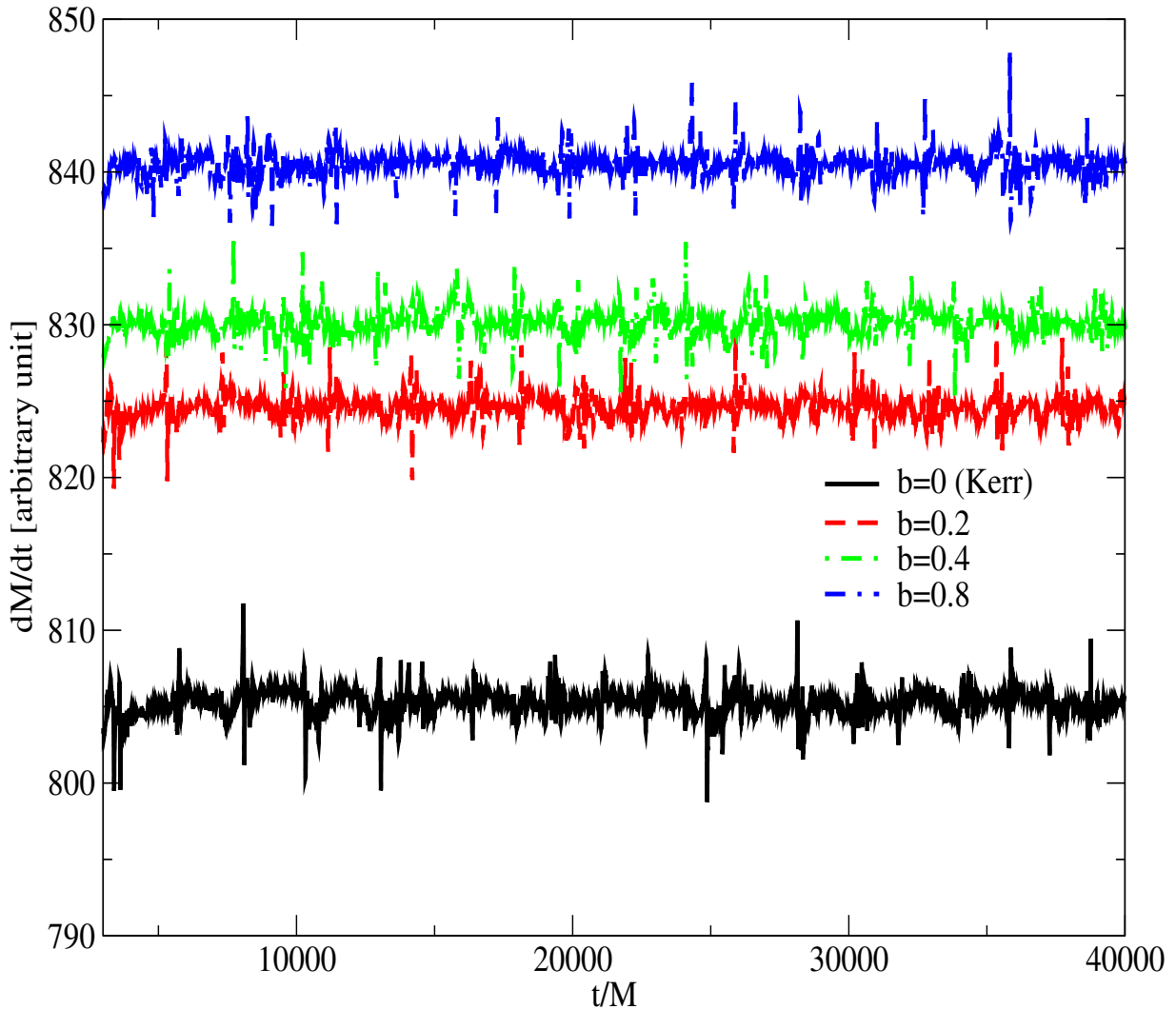}
\caption{
The time evolution of the mass accretion rate in the strong gravitational field around the QCBH with a spin parameter $a/M=0.9$ is shown for different values of the quantum correction parameter $b$. It is observed that as $b$ increases, the mass accretion rate also increases. Additionally, after the accretion disk reaches a steady state, it exhibits oscillatory behaviour around a certain value.
    }
\vspace{1cm}
\label{orhan_mass_acc}
\end{figure*}

To enable a meaningful comparison, we compute the mass accretion rate, which characterizes the behaviour of matter in the strong gravitational field near the ISCO. The mass accretion rate presented in Fig.\ref{orhan_mass_acc} is calculated at $r=2.3M$, and its time evolution is shown for different quantum correction parameter $b$ values. As in the rest of the paper, all the parameters in this figure are expressed in geometrized units.
As shown in Fig.\ref{orhan_mass_acc}, the parameter $b$ significantly alters the amount of matter falling towards the black hole. As $b$ increases, the amount of accreted matter shows a noticeable increase. The mass accretion rate obtained from the numerical simulations is consistent with the effective potential derived from the motion of the test particle, as shown in Figure 2. As is well known, a deeper effective potential allows more matter to accrete and fall toward the black hole. In Fig.\ref{fig_eff}, the depth of the effective potential increases with increasing $b$, and a similar trend is observed in the mass accretion rates obtained from the numerical simulations in Fig.\ref{orhan_mass_acc}. Therefore, the theoretical and numerical results agree well.

However, as also shown in Fig.\ref{orhan_mass_acc}, it exhibits a quasiperiodic behaviour after the accretion disk reaches a steady state. This behaviour indicates the formation of QPOs in the strong gravitational field. As also seen in Fig.\ref{orhan_mass_acc}, the increase in $b$ has caused a slight decrease in the amplitude of the oscillations. In this case, the resulting QPOs are expected to have lower amplitudes and potentially altered QPO frequencies. As a result, the quantum correction parameter $b$ has significantly impacted the physical phenomena that occur around QCBH. Since these effects lead to observable changes, this type of model—incorporating deviations from the Kerr black hole geometry—could potentially explain observational data that cannot be fully accounted for using the standard Kerr geometry alone.

As mentioned earlier, the QPOs seen in the mass accretion rate profiles shown in Fig.\ref{orhan_mass_acc} indicate the presence of QPO frequencies. To calculate the resulting QPO frequencies, we performed a Power Spectral Density (PSD) analysis using the mass accretion rate for both the Kerr black hole and the QCBH with $b=0.8$.

\begin{figure*}
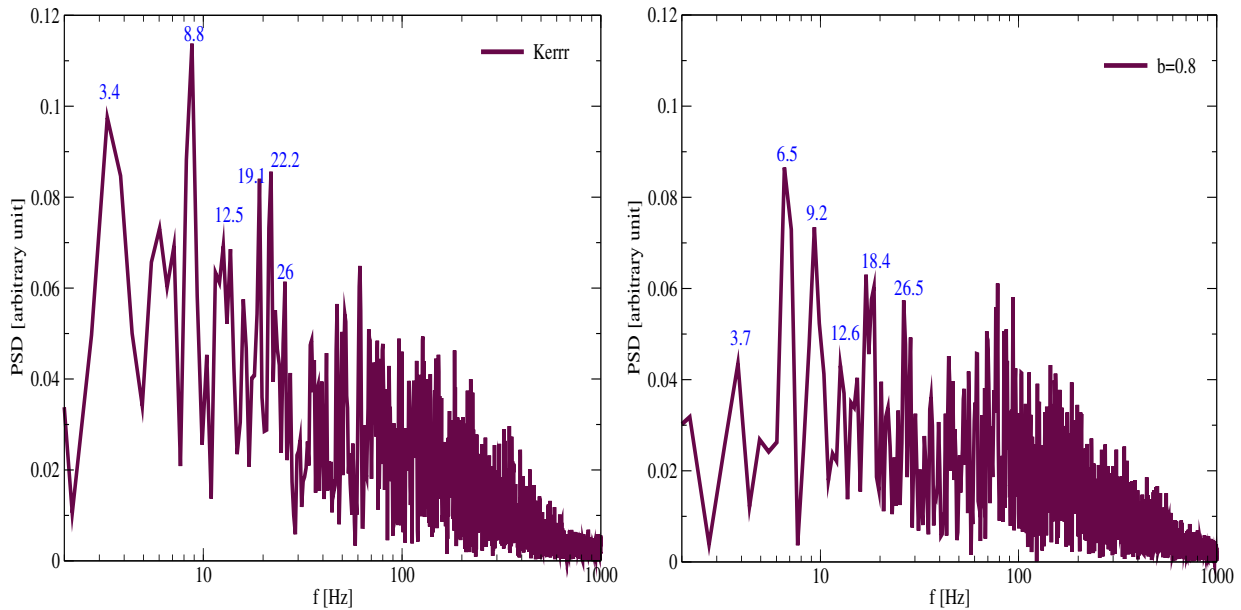

  \vspace{1cm}
  \center
  \psfig{file=PSDa09b0.eps,width=8.0cm,height=8.0cm} 
  \psfig{file=PSDa09b08.eps,width=8.0cm,height=8.0cm} 
\caption{
PSD profiles as a function of frequency for a black hole with  $M = 10\,M_\odot$ are shown for both the Kerr and QCBH cases. In the QCBH scenario, the quantum correction parameter is set to \( b = 0.8 \).  While QPO frequencies appear in both cases, the Kerr black hole exhibits a higher peak PSD amplitude, indicating stronger dominant oscillatory modes. In contrast, the QCBH case shows a redistribution of spectral power toward lower frequencies, resulting in a broader and more flattened PSD profile.} 
\vspace{1cm}
\label{orhan_QPOs}
\end{figure*}

The results of this analysis, presented in Fig.\ref{orhan_QPOs}, show that the PSD amplitudes in the Kerr case are significantly higher than those in the QCBH case. PSD amplitudes are a key factor affecting whether QPO frequencies can be observed. The lowest frequency peaks for the Kerr and QCBH models appear at 3.4 and 3.7 Hz, respectively. Previous studies and observational results indicate that these low frequencies arise due to spacetime frame-dragging effects caused by black hole spin. They are known as Lense-Thirring frequencies \citep{Donmez2025JHEAp}. These frequencies are nearly identical in both cases, indicating that the quantum correction parameter $b$ does not significantly alter this specific QPO component. This result is also supported by the analytical solution shown in Fig.\ref{fig_LT}, which is based on the motion of a test particle around the QCBH. Section \ref{orhan1} provides a detailed discussion regarding the observational, numerical, and theoretical implications of QPO frequencies that arise due to the Lense-Thirring effect.

In Fig.\ref{orhan_QPOs}, additional peaks are observed in both models that result from the trapping of pressure modes within the accretion disk and their non-linear couplings. However, the dominant QPO peak shifts from 8.8 Hz (Kerr) to 6.5 Hz (QCBH) when the quantum correction parameter $b$ is introduced. This shift arises entirely from modifying the gravitational potential around the QCBH by the parameter $b$. This effect is also shown for the test particle case in Fig. \ref {fig_eff}.

In conclusion, the gravitational potential, turbulence and instabilities around the QCBH are influenced by the quantum correction parameter $b$, which leads to modifying the spacetime structure in the strong gravity region. It causes a shift in the ISCO location and a change in the depth of the effective potential, both of which result in changes in the QPO frequencies. These changes are clearly demonstrated in both the test particle model and the numerical simulations. Moreover, they also affect the observability of QPOs. Lower QPO amplitudes and shifted frequencies in the QCBH model suggest that specific observed QPO signals in black hole X-ray binaries may not be fully explained by Kerr spacetime alone. The QCBH framework could provide new information on unexplained QPO behaviors, such as weak or shifted QPOs.

\section{QPO Formation via Lense-Thirring Precession: A Comparative Study of Test Particle and Accretion Disk Models} \label{orhan1}

Analyzing QPOs around black holes will not only help us understand the phenomena occurring in strong gravitational fields. Still, it will also enable us to reveal the physical properties of the black hole itself. Comparison of the QPO frequencies observed in X-ray data from specific sources with those obtained through theoretical and numerical calculations will significantly contribute to the literature. In this context, this study aims to contribute to the literature by examining the similarities and differences between the QPOs that arise due to the Lense-Thirring effect around QCBHs — calculated using test particles — and the QPOs obtained from accretion disk models based on both the Kerr metric and alternative gravity theories.

The particle model investigates the geodesic motion of a test particle around the QCBH. Therefore, the calculations do not include non-gravitational forces, such as pressure. From the motion of the test particle, the orbital, radial, and vertical frequencies are determined in the strong gravitational field region of the QCBH ($r < 10M$) are determined. Based on these frequencies, the variation of the precession frequency caused by the Lense-Thirring effect is analysed with respect to the black hole spin and quantum deformation parameters. On the other hand, in the disk model, plasma structures formed around black holes through BHL accretion are modelled in different gravity theories, and the resulting QPO behaviour is explored \cite{ Donmez2014MNRAS,Donmez2024Universe, Donmez2024MPLA, Donmez2025JHEAp}. Numerical analyses reveal how QPOs resulting from the Lense-Thirring effect are influenced by physical factors such as the black hole's spin parameter, the physical characteristics of the accretion mechanism, and the shock waves that form. In the disk model, the QPO frequencies obtained give the same value at any point within the strong gravitational field. In contrast, in the particle model, these frequencies vary depending on the radial coordinate $r$.

Theoretically, it is known that the QPOs calculated in the test particle model originate from modes trapped within the potential well near the black hole horizon. Of course, the potential well is influenced by the black hole spin parameter and the quantum deformation parameter. These factors lead to variations in QPO frequencies. In contrast, in the disk model, it is understood that these QPOs arise not only due to the black hole's spin parameter but also from the trapping of modes by shock waves formed on the disk — and even from their nonlinear couplings. The modes trapped in the disk model are identified as $g-$modes and $p-$modes. As seen in Figure 6, the periapsis precession frequency is affected by the black hole's spin and the quantum deformation parameter.
In contrast, as shown in Figure 7, it is demonstrated that the Lense-Thirring frequency is influenced solely by the black hole spin parameter. However, in the disk model, similar to the test particle model, the effect of the black hole spin parameter is visible (see Figures 3 and 7 in \cite{Donmez2025arXiv250316665D}, Figures 2 and 5 in \cite{DONMEZ2024PDU}, and Figure 16 in \cite{Donmez2024JCAP}). In addition, it has been numerically shown under different gravity theories that the shock waves occurring in the strong gravitational field, the nature of the accretion mechanism and its initial conditions, and the suppression or enhancement of modes by pressure forces influence the amplitude of QPOs arising from the Lense-Thirring effect, either increasing or decreasing it \cite{Donmez2024JCAP, Donmez2022Univ, Donmez2014MNRAS,Donmez2024Universe, Donmez2017MPLA}.

\begin{table}
\footnotesize
\caption{
Comparison of LFQPOs obtained from observational results of different sources with QPOs derived from numerical simulations of the accretion disk and analytically calculated test particle QPOs. Here, $a/M$ represents the spin parameter of the black hole, $\mathcal{M}$ denotes the Mach number of the accreting matter in the accretion disk scenario.}
 \label{QPO_table}
\begin{center}
\vspace*{-0.2cm}
  \begin{tabular}{ccccc}
    \hline
    \hline
    Source & Mass $(M_{\odot})$   & $a/M$ & LFQPOs (Hz) & $\mathcal{M}$  \\
    \hline
    \textit{Observation} &  & & & \\
    \hline
    $GRS 1915+105$  & $12–18$ & $0.98$     & $1-10$      &  --\\
     \cite{Strohmayer2001ApJ,Liu_2021,Majumder_2022,Motta2023}    & & & & \\
    $GRO J1655-40$  & $6.3$   & $0.3-0.9$   & $0.1-10$    &  --  \\
     \cite{Remillard1999ApJ,Motta2014MNRAS} & & & & \\
    $XTE J1550-564$ & $9.1$   & $0.34-0.55$   & $0.1-10$    &  -- \\
    \cite{Varniere_2018} & & & & \\
    $H1743-322$     & $8–10$  & $0.2-0.48$   & $0.1-10$         & --  \\
     \cite{2012ApJ...754L..23A} & & & & \\
      \hline
    \textit{Accretion Disk} \cite{Donmez2025arXiv250316665D} & & & & \\   
    \hline
    -- & $10$ & $0.9$ & No & $1$ \\
    -- & $10$ & $0.9$ & $6.6$ & $2$ \\   
    -- & $10$ & $0.9$ & $1.5$ & $3$ \\   
    -- & $10$ & $0.9$ & $7.4$ & $4$ \\   
    -- & $10$ & $0.5$ & $4$ & $2$ \\   
    -- & $10$ & $0.0$ & No & $2$ \\   
    \hline
    \textit{Test Particle} ($b=0.2$) & & & & \\   
    \hline
    -- & $10$ & $0.9$ & $5.1$ & -- \\ 
    -- & $10$ & $0.5$ & $3.02$ & -- \\     
    -- & $10$ & $0.0$ & No & -- \\     
    \hline
    \hline
  \end{tabular}u
\end{center}
\vspace*{-0.5cm}
\end{table}

It is well known that LFQPOs generally arise from the Lense-Thirring precession effect. This effect is entirely a consequence of spacetime being warped in the strong gravitational field near the event horizon of a spinning black hole. The LFQPO frequencies observed in different sources from various observational datasets, the frequencies we calculate based on the motion of a test particle in this study, and those obtained from our numerical solutions of the GRH using an accretion disk model in the strong gravity regime \cite{Donmez2025arXiv250316665D} are listed in Table \ref{QPO_table}. In this study, we find that the Lense-Thirring precession frequency derived analytically from the precessional motion of a test particle around the black hole is in strong agreement with the QPO frequencies resulting from our GRH-based numerical accretion disk simulations, which model the interaction between matter and the black hole in a strong gravitational field. Moreover, the theoretical and numerical results agree with the observational data. As observed in the analytically derived test particle models and the numerically modelled accretion disk, no LFQPOs are generated for a non-rotating black hole (i.e. for spin parameter $a=0$). This result serves as a key confirmation that the observed LFQPOs originate from Lense-Thirring precession. Since a non-spinning black hole does not induce spacetime frame dragging, these oscillations do not occur in such cases.
Furthermore, our accretion disk model shows that even in the presence of a rapidly spinning black hole, if the infalling matter exhibits purely sonic behaviour, LFQPOs are not generated. It is also observed that the LFQPO frequency varies depending on whether the accreting matter is subsonic or supersonic. This finding suggests that the behaviour of matter near the black hole, shaped by the curvature of spacetime, directly impacts the generation of QPOs. Notably, QPOs that might be predicted analytically from idealized test particle motion are not always observable, highlighting the importance of incorporating full fluid dynamics and relativistic effects in modelling.

In summary, both test particle and accretion disk models provide crucial insights into the nature of QPOs formed around black holes. These studies are essential for understanding the origin of QPOs and proposing physical mechanisms that explain the QPOs observed in the data. While the test particle model is crucial in understanding QPO behaviour and for testing relativistic orbital models, accretion disk models are valuable for uncovering the rich dynamical physical structures and exploring how they vary under Kerr and alternative gravity theories. With the accretion disk model, we reveal the fundamental modes and calculate the QPO frequencies that result from the nonlinear coupling these modes can produce. Therefore, such studies will shed light on interpreting future observational data.
\section{Conclusions} \label{s4}
This work comprehensively investigates epicyclic oscillations around rotating QCBHs, exploring how quantum corrections influence particle dynamics, accretion processes, and observable phenomena in the strong gravity regime. The key results of our study can be outlined as follows:

\begin{itemize}
    \item The quantum correction parameter $b$ and rotation parameter $a$ significantly modify the energy and angular momentum profiles of particles in stable circular orbits (Figs. \ref{figENG}-\ref{fig_ANG}). Compared to the classical Kerr solution, particles orbiting QCBHs exhibit higher energy and angular momentum values, with these differences becoming more pronounced near the event horizon.

    \item The effective potential analysis (Fig. \ref{fig_eff}) indicates that quantum corrections enhance the potential barrier, making circular orbits less stable compared to the Kerr case. This effect is particularly noticeable for larger values of $b$, indicating that quantum corrections impact accretion disk structure and stability (Fig. \ref{orhan_mass_acc}).

    \item The effective force experienced by test particles (Fig. \ref{figForce}) shows stronger attractive behaviour in QCBHs compared to Kerr black holes, with the force magnitude increasing with both $a$ and $b$ parameters, which has implications for particle capture rates and accretion processes.

\item Our calculation of fundamental oscillation frequencies (Fig. \ref{figFRQ}) demonstrates that quantum corrections shift the radial profiles of $\nu_r$, $\nu_\theta$, and $\nu_\phi$ closer to the black hole. The frequency ratios, particularly the 3:2 ratio often observed in QPOs, show measurable deviations from Kerr predictions that future high-precision X-ray timing observations could potentially constrain.

\item The precession effects (Figs. \ref{fig_periastron}-\ref{fig_LT}) exhibit distinct dependence on the quantum correction parameter. While the Lense-Thirring precession is primarily governed by the rotation parameter $a$, the periastron precession shows sensitivity to $a$ and $b$, providing an additional channel to probe quantum corrections.

\item  The radial ($\nu_r$), vertical ($\nu_\theta$), and orbital ($\nu_\phi$) frequencies (Fig.~\ref{figFRQ}) are influenced by $a$ and $b$. For non-rotating BHs ($a=0$), $\nu_r$ and $\nu_\theta$ coincide, but increasing $a$ or $b$ breaks this degeneracy, shifting $\nu_r$ to smaller radii---indicating tighter orbital confinement near the black hole.

    \item  Periastron precession ($\Omega_P$, Fig.~\ref{fig_periastron}) decreases with $a$, while Lense-Thirring precession ($\Omega_{LT}$, Fig.~\ref{fig_LT}) increases with $a$, highlighting how frame-dragging alters orbital planes in rotating spacetimes.

   \item  Through numerical simulations, it has been demonstrated that the plasma structure formed due to BHL accretion of matter around the QCBH generates QPOs by trapping pressure modes, which in turn excite various modes that give rise to QPO frequencies. The numerical results reveal the presence of fundamental pressure modes and additional modes formed through their nonlinear couplings. These modes have been numerically confirmed to depend on the quantum correction parameter $b$ strongly. Additionally, the simulations show the presence of mode frequencies arising from the Lense-Thirring effect. It has been confirmed—both in the accretion disk model and the test particle model—that this particular frequency changes only with the spin parameter of the black hole, and the influence of $b$ on it is negligibly small.\
   
    \item The parameter $b$ has a negligible impact on $\Omega_{LT}$, suggesting that quantum corrections predominantly modify the radial dynamics rather than the spacetime rotation effects, which remain dominated by the spin parameter $a$.
    At the same time, it has been demonstrated that the Lense-Thirring frequency is consistent with the LFQPOs observed from some sources and obtained through numerical solutions of the GRH equations in strong gravitational fields, which, in turn, reinforces the validity of the studies conducted here.
    
\end{itemize}

These results have several important implications for astrophysical observations and theoretical studies: The modifications to orbital frequencies and precession effects suggest that quantum corrections could potentially explain certain anomalies in observed QPO spectra from microquasars and AGNs. Our frequency calculations provide concrete predictions that can be compared with data from current and future X-ray timing missions. The enhanced effective potential and force effects indicate that quantum corrections may influence the thermodynamic properties of the accretion disk and the emission properties. It could be relevant for interpreting the spectra of black holes where quantum gravity effects might be significant, such as primordial or microscopic holes in certain theories. Our work provides a framework for testing quantum gravity models through astrophysical observations. The measurable differences from Kerr predictions, particularly in the strong-field regime near the ISCO, offer potential observational signatures for quantum corrections.

Several directions for future research emerge from this work. One can extend our analysis to include charged particle dynamics and electromagnetic field effects, which would be relevant for magnetized accretion disks. Investigating resonance phenomena between different oscillation modes could provide enhanced observational signatures of quantum corrections. More detailed models were developed to connect our particle dynamics results to actual QPO observations, including radiative transfer calculations through quantum-corrected spacetimes.
Application of our framework to specific quantum gravity models to derive constraints on fundamental parameters from astrophysical data.

In conclusion, our study demonstrates that quantum corrections to black hole spacetimes can have observable consequences for particle dynamics and oscillation frequencies. Although current observational precision may not yet allow for definitive detection of these effects, next-generation X-ray timing instruments and gravitational wave detectors could provide the necessary sensitivity to test these predictions. The framework developed here is an essential step toward bridging quantum gravity theories with astrophysical observations of black holes.

\section*{Acknowledgments}
All numerical simulations were performed using the Phoenix High
Performance Computing facility at the American University of the Middle East (AUM), Kuwait.  S.G.G. would like to thank ANRF, India for project No. ANRF/2021/005771. \\

\bibliographystyle{apsrev4-1}  
\bibliography{referenceR}

\end{document}